\documentclass[final,5p,times,twocolumn]{elsarticle}
\usepackage{url}

\usepackage{amssymb}
\usepackage{amsmath}
\usepackage{bbm}

\usepackage{tikz}
\usetikzlibrary{snakes}

\begin{document}

\begin{frontmatter}



\title{Two-point tree-level string amplitudes as AdS transition amplitudes}


\author[PS]{Pongwit Srisangyingcharoen}
\ead{pongwits@nu.ac.th}
\author[JP]{Jongruk Pukdee}
\affiliation[PS]{organization={The Institute for Fundamental Study, Naresuan University},
            city={Phisanulok},
            postcode={65000}, 
            country={Thailand}}

\affiliation[JP]{organization={Department of Physics, School of Science, Nagoya University},
            city={Nagoya},
            postcode={}, 
            country={Japan}}

\begin{abstract}
We compute the two-point open string and closed string amplitudes at tree level and show that, in a 't Hooft-like limit, they take a form structurally analogous to boundary-to-boundary transition amplitudes of a scalar field in Euclidean AdS space. Interestingly, while both amplitudes yield equivalent expressions, the associated ’t Hooft couplings are defined by different worldsheet curvatures—geodesic for the disk and Gaussian for the sphere—suggesting a possible geometric aspect of open/closed string duality.
\end{abstract}




\end{frontmatter}


\section{Introduction}\label{sec1}
An essential framework for exploring the interplay between gravity and quantum field theory is the AdS/CFT correspondence \cite{Maldacena:1997re}. The idea that quantum gravity may be fundamentally holographic was first inspired by the thermodynamic properties of black holes, particularly the Bekenstein–Hawking entropy formula \cite{PhysRevD.7.2333,Hawking:1974rv}. The AdS/CFT correspondence offers the first concrete realization of this holographic principle within string theory \cite{Maldacena:1997re,Witten:1998qj,Gubser:1998bc}. Beyond its conceptual significance, the correspondence has become a powerful computational tool for probing strongly coupled, non-perturbative regimes of quantum field theories. See \cite{DHoker:2002nbb, Nastase:2007kj,Aharony:1999ti} for a comprehensive review.

A particularly intriguing perspective on this duality was presented by Gopakumar \cite{Gopakumar:2003ns}. In his work, he showed that one-loop planar correlation functions in free field theories can be reorganized into a form that mimics field amplitudes in AdS space. This was accomplished by expressing Feynman diagrams as integrals over Schwinger parameters, which naturally matched with the bulk-to-boundary propagators of AdS scalar fields. 

Motivated by this insight, we investigate whether similar structures arise directly from open string amplitudes in flat space, without appealing to a field theory limit. In particular, we analyze the two-point open string amplitude at tree level and study whether, in a suitable limit, it can be cast into a form structurally resembling an AdS boundary-to-boundary transition amplitude. Our goal is to understand if and how open string correlators, under appropriate parametrization and scaling, reveal an intrinsic connection to propagators in AdS space—potentially offering a new perspective on the emergence of holographic dualities.

\section{AdS Scalar propagator and transition amplitude}\label{sec2}
In this section, we would like to exhibit the bulk-to-boundary propagators in $\text{AdS}_{d+1}$ in a Schwinger representation for a massive scalar field. It is convenient to work in the Poincar\'e coordinates of Euclidean $\text{AdS}_{d+1}$,
\begin{equation}
    ds^2=\frac{d\vec{z}^2+dz_0^2}{z_0^2}.
\end{equation}
Let us consider the Euclidean action
\begin{equation}
    S[\phi]=\frac{1}{2}\int d^dz dz_0 \sqrt{g}\left(g^{MN}\partial_M\phi(z)\partial_N\phi(z)+m^2\phi(z) \right)
\end{equation} 
where $\phi(z)=\phi(z_0,\vec{z})$. A bulk-to-boundary propagator $K(z_0,\vec{z};\vec{z}')$ can be obtained from the wave equation
\begin{equation}
    \left(z_0^{d+1}\frac{\partial}{\partial z_0}\left(z_0^{-d+1}\frac{\partial}{\partial z_0}  \right)+z_0^2\Box-m^2\right)K=0. \label{bulktoboundary prop}
\end{equation}
where $\Box$ is $d$-dimensional Laplacian operator in direction $\vec{z}$. It is well-known that the expression for bulk-to-boundary propagator takes the form 
\begin{equation}
    K(z_0,\vec{z};\vec{z}')=\frac{\Gamma(\Delta)}{\pi^{d/2}\Gamma(\Delta-\frac{d}{2})}\left( \frac{z_0}{z_0^2+(\vec{z}-\vec{z}')^2}\right)^\Delta
\end{equation}
where $\Delta=\frac{1}{2}\left(d+\sqrt{d^2+4m^2}\right)$ \cite{Witten:1998qj}. The AdS bulk field $\phi(z)$ is related to the boundary data by
\begin{equation}
    \phi(z_0,\vec{z})=\int d^d\vec{x} \ K(z_0,\vec{z};\vec{x})\phi_0(\vec{x}).
\end{equation}

Using a Schwinger parametrization, one obtains
\begin{equation}
    K(z_0,\vec{z};\vec{z}')=\frac{1}{z_0^\Delta\pi^{d/2}\Gamma(\Delta-\frac{d}{2})}\int_0^\infty d\rho \ \rho^{\Delta-1}e^{-\rho}e^{-\frac{(\vec{z}-\vec{z}')^2}{z_0^2}\rho}
\end{equation}
by which we can utilize the heat kernel,
\begin{equation}
    \langle x|e^{\tau\Box}|y\rangle=\frac{1}{(4\pi \tau)^{d/2}}e^{-(x-y)^2/4\tau}
\end{equation}
to rewrite
\begin{equation}
    K(t,\vec{z};\vec{z}')=\frac{t^{(d-\Delta)/2}}{\Gamma(\Delta-\frac{d}{2})}\int_0^\infty d\rho \ \rho^{\Delta-\frac{d}{2}-1} e^{-\rho} \langle \vec{z}|e^{\frac{t}{4\rho}\Box}|\vec{z}'\rangle
\end{equation}
where $t=z_0^2$. We would write $\vec{z}$ as $z$ for later convenience.

We can then construct a boundary-to-boundary propagator through
\begin{align}
   \Gamma(x,y)=&\frac{1}{2}\int_0^\infty \frac{dt}{t^{\frac{d}{2}+1}}\int d^dz \ K(t,z;y)K(t,z;x)    \nonumber \\
   =&\frac{1}{2(\Gamma(\Delta-d/2))^2}\int_0^\infty \frac{dt}{t^{\Delta+1-\frac{d}{2}} } \int d^dz\int_0^\infty d\rho_1 \int_0^\infty d\rho_2 \nonumber \\
   &\times \ (\rho_1\rho_2)^{\Delta-\frac{d}{2}-1} e^{-(\rho_1+\rho_2)}\langle y|e^{\frac{t}{4\rho_2}\Box}|z\rangle\langle z|e^{\frac{t}{4\rho_1}\Box}|x\rangle. \label{2-pt ads prop}
\end{align}
 While equation (\ref{2-pt ads prop}) describes the standard boundary-to-boundary propagator constructed from two identical bulk-to-boundary propagators of weight $\Delta$, one may consider a generalized form involving distinct weights $\Delta_1$ and $\Delta_2$. This leads to
\begin{align}
    &\Gamma(x,\Delta_1; y,\Delta_2) = \frac{1}{2\Gamma(\Delta_1-\frac{d}{2})\Gamma(\Delta_2-\frac{d}{2})}\int_0^\infty \frac{dt}{t^{1+(\Delta_1+\Delta_2-d)/2 }} \int d^dz \nonumber \\
    &\times \int_0^\infty d\rho_1 \int_0^\infty d\rho_2 \ \rho_1^{\frac{\Delta_1-d}{2}-1}\rho_2^{\frac{\Delta_2-d}{2}-1}e^{-(\rho_1+\rho_2)}\langle y|e^{\frac{t}{4\rho_2}\Box}|z\rangle\langle z|e^{\frac{t}{4\rho_1}\Box}|x\rangle \label{trans amp}
\end{align}
which can be interpreted as a transition amplitude between scalar fields of different weights. 

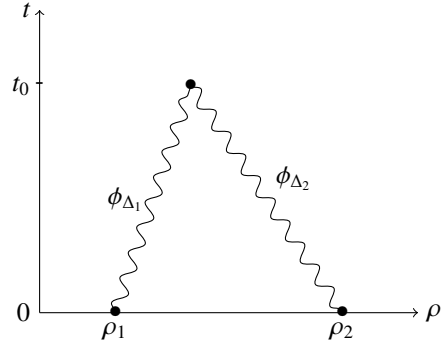
\begin{figure}
\centering
\begin{tikzpicture}
\draw[->] (0,0)--(5,0);
\draw[->] (0,0)--(0,4);
\node[left] at (0,0) {0};
\node[right] at (5,0) {$\rho$};
\node[left] at (0,4) {$t$};
\node at (1,0) {\textbullet};
\node[below] at (1,0) {$\rho_1$};
\node at (4,0) {\textbullet};
\node[below] at (4,0) {$\rho_2$};
\node at (2,3) {\textbullet};
\node at (0,3) {-};
\node[left] at (0,3) {$t_0$};
\draw [snake=snake] (1,0)--(2,3)--(4,0);
\node[left] at (1.5,1.5) {$\phi_{\Delta_1}$};
\node[above right] at (3,1.5) {$\phi_{\Delta_2}$};
\end{tikzpicture}
    \caption{A propagation of the scalar field in configuration space from the boundary to the boundary through an interaction point located in the bulk.}
    \label{prop}
\end{figure}

For future reference, let us denote 
\begin{align}
    2\Gamma(\Delta_1-d/2)&\Gamma(\Delta_2-d/2)\Gamma(x,\Delta_1; y,\Delta_2)\nonumber \\
    &=\Gamma_1(x,\Delta_1; y,\Delta_2)+\Gamma_2(x,\Delta_1; y,\Delta_2) 
\end{align}
where $\Gamma_1$ and $\Gamma_2$ refer to (\ref{trans amp}) evaluated in different integration domains: $\rho_1<\rho_2$ for $\Gamma_1$ and $\rho_1>\rho_2$ for $\Gamma_2$ by which we refer to this integration region as $\mathcal{I}_1$ and $\mathcal{I}_2$ respectively. Furthermore, we would like to introduce $t$-dependent transition amplitude as 
\begin{align}
    \Gamma_i(x,\Delta_1; y,\Delta_2;t)=&\int d^dz \int_{\mathcal{I}_i}d\rho_1 d\rho_2 \ \rho_1^{\frac{\Delta_1-d}{2}-1}\rho_2^{\frac{\Delta_2-d}{2}-1} \nonumber \\
    &\times e^{-(\rho_1+\rho_2)}\langle y|e^{\frac{t}{4\rho_2}\Box}|z\rangle\langle z|e^{\frac{t}{4\rho_1}\Box}|x\rangle. \label{t-dep trans amp}
\end{align}
Intuitively, this function could be interpreted as a partial transition amplitude which is in the same sense as color-ordered amplitude in QFT or in the context of open string amplitude, where the vertex operators of the states are inserted along the worldsheet boundary in an ordering way. This can be illustrated in figure(\ref{prop}) as a propagation of the scalar field $\phi_{\Delta_1}(\vec{x},0)$ from the point on the boundary (parameterized by $\rho_1$) to the point $(\vec{z},t_0)$ in the bulk. At this point, the field transitions into the field with different weight $\Delta_2$ and then propagates to the boundary point $(\vec{y},0)$ parameterized by $\rho_2$.

\section{Two-point Open String Amplitudes}\label{sec3}

In string theory, amplitudes are computed as correlation functions of vertex operators on Riemann surfaces. For open strings, these operators are inserted along the boundary of a disk, which is typically mapped conformally to the upper half-plane. It was widely believed that two-point string amplitudes vanish due to residual worldsheet symmetries after fixing two insertion points which have infinite volume. Until quite recently, it is pointed out that  the resulting divergence cancels the $\delta(0)$ arising from momentum conservation in the path integral formulation \cite{Erbin:2019uiz}. Since then, few follow-up works have explored this issue using alternative approaches \cite{Giribet:2023gub,SEKI2020135078,Kashyap:2020tgx,Kishimoto:2024aig}.

In this paper, we would like to compute a two-point open string amplitude with the aim of recasting its form to resemble that of a boundary-to-boundary AdS transition amplitude (\ref{trans amp}). The motivation is to explore manifestations of the AdS/CFT correspondence, under the expectation that the string amplitude—interpreted from the CFT side—would exhibit a structure analogous to the AdS amplitude.

An expression for the two-point open string amplitude at tree level is given by
\begin{equation}
    A(k_1,k_2)=\frac{1}{Z}\int DXDg \ \mathcal{V}_1(k_1)\mathcal{V}_2(k_2) e^{-S_P[X,g]}
\end{equation}
where the Polyakov action is
\begin{equation}
    S_P[X,g]=\frac{1}{4\pi \alpha'}\int_\Sigma d^2\sigma \sqrt{g} g^{ab}\partial_a X^\mu\partial_b X^\nu\eta_{\mu\nu}. \label{Pol}
\end{equation}
$\mathcal{V}_i(k_i)$ is an integrated vertex operator taking the form
\begin{equation}
    \mathcal{V}_i(k_i)=g_o\int_{\partial\Sigma} dx_i \ V_i(x_i,k_i)
\end{equation}
with an open-string coupling constant $g_o$. The denominator $Z$ is a partition function given by
\begin{equation}
    \int DXDg \ e^{-S_P[x,g]}.
\end{equation}
Using integration by parts, the action becomes
\begin{equation}
        \frac{1}{4\pi \alpha'}\int_\Sigma d^2\sigma \left(\sqrt{g} X^\mu \nabla^2 X_\mu -\partial_a\left( \sqrt{g} g^{ab} X^\mu\partial_b X_\mu\right)\right) \label{action}
\end{equation}
where 
\begin{equation}
    \nabla^2=\frac{1}{\sqrt{g}}\partial_a\left( \sqrt{g} g^{ab}\partial_b\right)
\end{equation}
is the Laplace–Beltrami operator. The second term represents a boundary contribution, which vanishes under Neumann boundary conditions, i.e., $\frac{\partial X}{\partial n}\biggr\rvert_{\partial\Sigma}=0$. 

For simplicity, we fix the worldsheet metric to $g^{ab}=\delta^{ab}$, which leaves the residual gauge symmetry group as PSL$(2,\mathbb{R})$. To simplify the computation, we restrict our attention to amplitudes involving the simplest form of vertex operators:
\begin{equation}
    V_j(x_j,k_j)=e^{ik_j\cdot X(x_j)}. \label{tach ver}
\end{equation}
with a Euclidean on-shell condition $k_i^2=-\frac{1}{\alpha'}$ (We use mostly minus Minkowski signature, accordingly, the Euclidean mass-shell condition reads $k^2=-k_{M}^2=m^2$ upon a Wick rotation). 

Expanding $X(\sigma)=\bar{X}(\sigma)+x$ where $x$ is a zero mode regarding the Laplacian operator, the measure reads $D\bar{X} d^dx$. Obviously the integration from the zero mode $x$ provides a momentum conservation, 
\begin{equation}
    (2\pi)^d\delta^{d}(k_1+k_2).
\end{equation}
What remains is a Gaussian integral,
\begin{align}
    \int D\bar{X} \exp\Bigg(&\int_\Sigma d^2\sigma  \Bigg[
        \frac{\sqrt{g}}{4\pi\alpha'} X^\mu \nabla^2 X_\mu \notag \\
        &\quad + i\sum_{i=1}^2 \delta^2(\sigma - x_i) k_i \cdot X(\sigma)
    \Bigg] \Bigg).
\end{align}
After performing a Gaussian integration, one would obtain
\begin{equation}
    \mathcal{N}\exp{\left( \pi\alpha' k_1\cdot k_2 G(x_1-x_2)\right)}
\end{equation}
where $G(x_1,x_2)$ is a Green's function satisfying
\begin{equation}
    \nabla^2G(x,y)=\frac{1}{\sqrt{g}}\delta^2(x-y)-\frac{1}{\text{Vol}} \label{Green function equation}
\end{equation}
with the Neumann boundary condition
\begin{equation}
    \frac{\partial G(x,y)}{\partial n_x}\Biggr|_{\partial\Sigma}=c \quad \text{(constant)}.
\end{equation}
Note that the factor $\mathcal{N}$ which involves the determinant of the Laplacian operator is subject to cancellation by the factor $Z$ in the denominator. The term $\frac{1}{\text{Vol}}$, where Vol is the volume of the compact space, is necessary to make the Green’s function solvable in the sense that it makes the Green's identity hold
\begin{align}
    &\int_\Sigma \left(X^\mu(x) \nabla^2G(x,y)-G(x,y)\nabla^2X^\mu(x)\right) dV_y \nonumber \\
    &=\oint_{\partial\Sigma}\left(X^\mu(y)\frac{\partial G(x,y)}{\partial \mathbf{n}_y}-G(x,y)\frac{\partial X^\mu(y)}{\partial \mathbf{n}_y} \right) dS_y
\end{align}
with vanishing boundary conditions, $\frac{\partial G(x,y)}{\partial \mathbf{n}_y}\bigg\vert_{\partial\Sigma}=\frac{\partial X^\mu(y)}{\partial \mathbf{n}_y}\bigg\vert_{\partial\Sigma}=0$.

Also, we neglect the exponents including the self-contracted terms like $k_i^2G(0)$, for $i=1,2$ as we need to avoid the self-contraction of operators within the normal-ordering. Remember that the two-point correlation function in state-operator formalism reads
\begin{equation}
    \sim\langle 0|:V_1(k_1)::V_2(k_2):|0\rangle.
\end{equation}


Let $\Sigma$ be a disk with radius $R$ on which the surface is parameterized by the usual polar coordinates $(r,\theta)$. By a technique of image charges, an expression for the Neumann Green's function is
\begin{align}
    G(r,\theta;r',\theta')=&\frac{1}{4\pi}\ln\Bigg| r^2+r'^2-2rr'\cos(\theta-\theta')\Bigg|\nonumber \\
    +&\frac{1}{4\pi}\ln\Bigg|r^2+\left(\frac{R^2}{r'}\right)^2-2r\left(\frac{R^2}{r'}\right)\cos(\theta-\theta')\Bigg|-\frac{r^2}{4\pi R^2}
\end{align}
which is subject to
\begin{equation}
    \frac{\partial G(r,\theta;r',\theta')}{\partial r}\Biggr|_{r=R}=0. \label{Neumann}
\end{equation}

At the boundary, the Green's function becomes
\begin{equation}
    G(\theta_1,\theta_2)=\frac{1}{\pi}\ln\Bigg| 2R\sin\left(\frac{\theta_1-\theta_2}{2}\right)\Bigg|-\frac{1}{4\pi}.
\end{equation}
Accordingly, the amplitude takes the form 
\begin{align}
    \mathcal{A}(k_1,k_2)=&(2\pi)^d\delta^d(k_1+k_2)g_o^2 \ e^{\alpha' k_1\cdot k_2/4 }\nonumber \\
    &\int_0^{2\pi}d\theta_1 \int_0^{2\pi}d\theta_2\Bigg|2R\sin\left(\frac{\theta_1-\theta_2}{2}\right) \Bigg|^{-\alpha'k_1\cdot k_2}.
\end{align}
We then change the coordinates 
\begin{equation}
    \theta^+=\frac{\theta_1+\theta_2}{2}, \qquad \theta^-=\frac{\theta_1-\theta_2}{2}.
\end{equation}
Therefore, we obtain
\begin{align}
    \mathcal{A}(k_1,k_2)=&2(2\pi)^{d+1}\delta^d(k_1+k_2)g_o^2 e^{\frac{\alpha'}{4} k^2}\int_{-\pi}^\pi d\theta^- \Big|2R\sin\left(\theta^-\right) \Big|^{-\alpha'k^2} \nonumber \\
    =&8g_o^2e^{-\frac{1}{4}}(2\pi)^{d+1}\delta^d(k_1+k_2)R\int_{0}^\pi d\theta^- \Big(\sin\left(\theta^-\right) \Big)^{-\alpha'k^2} \label{open amp}
\end{align}
where we applied the requirement $k_1=-k_2=k$ from the conservation of momentum together with the mass-shell condition, i.e. $k^2=-\frac{1}{\alpha'}$. Notice that we keep the $k^2$ untouched in the last line, expecting this to relate to the heat kernel presented in the AdS propagator.

We then re-express $\left(\sin(\theta^-)\right)^{-\alpha'k^2}$ as $\Im(e^{-i\alpha'k^2\theta^-})$ and rescale the variable $\theta^-=\theta/R$. Note that the expression is hold at the mass-shell level. Therefore, the amplitude takes the form
\begin{equation}
    8g_o^2(2\pi)^{d+1}\delta^d(k_1+k_2)\int_{0}^{\pi R} d\theta \ \Im(e^{-i\frac{\alpha'}{R}k^2\theta}) \label{2pi str}
\end{equation}
Note that we absorb the factor $e^{-\frac{1}{4}}$ into the string coupling $g_0$ for convenience. We observe that by taking the limit $R\rightarrow \infty$ while keeping $\alpha'/R$ fixed, it is possible to eliminate the imaginary operation through a Wick rotation: $\theta\rightarrow -i\theta$, turning the expression (\ref{2pi str}) into
\begin{equation}
-8g_o^2(2\pi)^{d+1}\delta^d(k_1+k_2)\int_0^{\infty}d\theta \ e^{-\lambda k^2\theta} \label{2pt str2}
\end{equation}
where $\lambda=\alpha'/R$ is held constant. This limit is reminiscent of the t'Hooft limit in large $N$ gauge theories \cite{tHooft:1973alw}. As a consequence of $\lambda$ being constant, our string theory is tensionless ($\alpha'\rightarrow\infty$). This regime has been investigated in several studies \cite{Isberg:1993av,Gross:1988ue,Bonelli:2003kh, Lizzi:1986nv}, including its potential connections to gauge theories \cite{Edwards:2014cga,Edwards:2014xfa,Srisangyingcharoen:2022ixt}.

To mimic the form of the AdS transition amplitude (\ref{trans amp}), we insert an identity
\begin{equation}
    1=\frac{1}{\Gamma(n+1)}\int_0^\infty du \ u^{n}e^{-u}
\end{equation}
with an arbitrary number $n$ into (\ref{2pt str2}). We then change the coordinate $\theta=\frac{1}{v}$ and transform the coordinates $(u,v)\rightarrow(\rho_1,\rho_2)$ via 
\begin{equation}
    u=\rho_1+\rho_2 \quad \text{and} \quad \frac{1}{v}=\frac{1}{\rho_1}+\frac{1}{\rho_2}.
\end{equation}
to give
\begin{align}
    -8g_o^2(2\pi)^{d+1}&\frac{\delta^d(k_1+k_2)}{\Gamma(n+1)}\int_0^\infty d\rho_1\int_0^\infty d\rho_2 \ \Bigg| \frac{\rho_1-\rho_2}{\rho_1+\rho_2}\Bigg| \nonumber \\
   \times & \left( \frac{1}{\rho_1}+\frac{1}{\rho_2}\right)^2(\rho_1+\rho_2)^n e^{-(\rho_1+\rho_2)}e^{-\lambda k^2\left( \frac{1}{\rho_1}+\frac{1}{\rho_2}\right)}. \label{2pt str3}
\end{align}
Now, we can perceive $k^\mu$ as an eigenvalue of the operator $\hat{k}$ acting on $|k\rangle$. This allows us to write
\begin{equation}
    (2\pi)^d\delta^d(k_1+k_2)e^{-\lambda k^2\left( \frac{1}{\rho_1}+\frac{1}{\rho_2}\right)}=\langle k_2|e^{-\lambda \hat{k}^2\left( \frac{1}{\rho_1}+\frac{1}{\rho_2}\right)}|k_1\rangle.
\end{equation}
where 
\begin{equation}
    \langle k_2|k_1\rangle=(2\pi)^d\delta^d(k_1+k_2).
\end{equation}
A Fourier transform is performed to turn the two-point amplitude (\ref{2pt str3}) into that in position space such that the amplitude becomes
\begin{align}
    &\mathcal{A}(x,y)=\int \frac{d^dk_1}{(2\pi)^d}\frac{d^dk_2}{(2\pi)^d}\mathcal{A}(k_1,k_2)e^{-i(k_1\cdot x-k_2\cdot y)} \nonumber \\
    &=\frac{-16g_o^2\pi}{\Gamma(n+1)}\int \frac{d^dk_1}{(2\pi)^d}\frac{d^dk_2}{(2\pi)^d}\int_0^\infty d\rho_1\int_0^\infty d\rho_2 \ \mathcal{F}(\rho_1,\rho_2) \nonumber \\
    &\hphantom{=}\times e^{-(\rho_1+\rho_2)}\langle y|k_2\rangle\langle k_2|e^{-\lambda \hat{k}^2\left( \frac{1}{\rho_1}+\frac{1}{\rho_2}\right)}|k_1\rangle\langle k_1|x\rangle
\end{align}
where 
\begin{equation}
    \mathcal{F}(\rho_1,\rho_2)=\Bigg| \frac{\rho_1-\rho_2}{\rho_1+\rho_2}\Bigg|\left( \frac{1}{\rho_1}+\frac{1}{\rho_2}\right)^2(\rho_1+\rho_2)^n
\end{equation}
and $\langle x|k\rangle=e^{ik\cdot x}$. The integration over $d^dk_i$, $i=1,2$ forms a completeness relation by which we can integrate out to obtain
\begin{align}
    \mathcal{A}(x,y)=&\frac{-16g_o^2\pi}{\Gamma(n+1)}\int_0^\infty d\rho_1\int_0^\infty d\rho_2 \int d^dz \ \mathcal{F}(\rho_1,\rho_2) \nonumber \\
    &\times  e^{-(\rho_1+\rho_2)}\langle y|e^{\frac{\lambda}{\rho_1}\Box}|z\rangle\langle z|e^{\frac{\lambda}{\rho_2}\Box}|x\rangle \label{2pt str4}
\end{align}
where we insert the identity, $\mathbbm{1}=\int d^dz \  |z\rangle\langle z|$ into the equation and recast $\hat{k}^2=-\Box$ in position space.

The expression (\ref{2pt str4}) is similar to the AdS transition amplitude (\ref{trans amp}) where the parameter $\lambda$ plays the same role as $t$. However, we still need to cope with a mismatch between the integrand, i.e. $\mathcal{F}(\rho_1,\rho_2)$ and $\rho_1^{\frac{\Delta_1-d}{2}-1}\rho_2^{\frac{\Delta_2-d}{2}-1}$. Obviously, we can rewrite
\begin{equation}
    \mathcal{F}(\rho_1,\rho_2)=\sum_{s=0}^{\infty}\binom{n+1}{s} \ \rho_1^{n-s-1}\rho_2^{s-2}|\rho_1-\rho_2|.
\end{equation}
Accordingly, the two-point amplitude can be rewritten as a combination of AdS transition amplitudes,
\begin{align}
    \mathcal{A}(x,y)=& \frac{-16g_o^2\pi}{\Gamma(n+1)}\sum_{s=0}^{\infty}\binom{n+1}{s} \sum_{j=1}^2 \ (-1)^{j}\nonumber \\ &\times\Bigg[\Gamma_j(x,\widetilde\Delta_1+2;y,\widetilde\Delta_2-2;4\lambda) 
    -\Gamma_j(x,\widetilde\Delta_1;y,\widetilde\Delta_2;4\lambda )\Bigg] \label{relation string and transition amp}
\end{align}
where 
\begin{equation}
    \widetilde\Delta_1=2(n-s)+d \quad \text{and} \quad \widetilde\Delta_2=2s+d.
\end{equation}
The expression for $\Gamma_j(x,\alpha;y,\beta;t)$ is shown in (\ref{t-dep trans amp}). Note that the summation is finite if $n$ is a non-negative integer.

\section{Two-point Closed String Amplitudes}
In this section, we would like to deliver a similar calculation to the previous section, but for the closed string states whose vertex operators are in the form
\begin{equation}
    \mathcal{V}_i(k_i)=g_c\int_\Sigma d^2x_i V_i(x_i,k_i).
\end{equation}
The factor $g_c$ is a closed-string coupling constant. We will take the simplest form of the vertex operators shown (\ref{tach ver}) with the Euclidean on-shell conditions, $k_i^2=-\frac{4}{\alpha'}$.

We can then repeat with a similar computation to obtain
\begin{align}
    \mathcal{A}(k_1,k_2)=&(2\pi)^d\delta^d(k_1+k_2)g_c^2 \nonumber \\
    &\times\int_\Sigma d^2x_1d^2x_2 \ \exp\left({\pi\alpha' k_1\cdot k_2 G(x_1-x_2)}\right)
\end{align}
where $G(x_1,x_2)$ is a Green's function that obeys (\ref{Green function equation}).

At tree-level, we consider a worldsheet being a sphere $S^2$ of radius $R$. It is known that the Laplacian Green's function for a unit sphere $S^2$ takes the form
\begin{align}
    G(\theta_1,\phi_1;\theta_2,\phi_2)=-\frac{1}{4\pi}\ln\left| 1-\cos\gamma\right|
\end{align}
where 
\begin{equation}
   \cos\gamma= \cos\theta_1\cos\theta_2+\sin\theta_1\sin\theta_2\cos(\phi_1-\phi_2)
\end{equation}
\cite{Kimura1987,Kimura1999}. The Green's function satisfies
\begin{equation}
    \nabla^2_{S^2}G(\gamma)=\frac{1}{2\pi}\delta(1-\cos\gamma)-\frac{1}{4\pi}.
\end{equation}
As the Laplacian for the radius-$R$ sphere is $\nabla^2_R=\frac{1}{R^2}\nabla^2_{S^2}$, the Green's function of that surface is 
\begin{equation}
    G_R(\gamma)=-\frac{1}{4\pi}\ln\left|R^2(1-\cos\gamma)\right|.
\end{equation}

Accordingly, one can write 
\begin{align}
    \mathcal{A}(k_1,k_2)=&(2\pi^2)(2\pi)^d\delta^d(k_1+k_2)g_c^2\nonumber \\
    &\times\int_0^\pi d\theta_1 \exp\left(-\frac{\alpha'}{4}k_1\cdot k_2  \ln\left|(1-\cos\theta_1)R^2\right| \right).
\end{align}
To obtain the above expression, we utilize a spherical symmetry which allows us to fix a point $(\theta_2,\phi_2)$ at the north pole. Therefore,
\begin{align}
    \mathcal{A}(k_1,k_2)=\pi^2(2\pi)^d\delta^d(k_1+k_2)g_c^2 \frac{1}{R^2}\int_0^\pi d\theta_1 \left(\sin\left(\frac{\theta_1}{2}\right)\right)^{\frac{\alpha'}{2} k^2}.
\end{align}
where the conservation of momentum and the mass-shell condition, $k^2=-\frac{4}{\alpha'}$, were applied. Let's rescale the variable $\theta_1= 2\theta R^2$, the amplitude becomes
\begin{align}
    2\pi^2(2\pi)^d\delta^d(k_1+k_2)g_c^2 \int_0^{\frac{\pi}{2R^2}} d\theta  \left(\sin{(2R^2\theta)}\right)^{\frac{\alpha'}{2} k^2}. \label{closed amp}
\end{align}
To deal with the integral term, we use the negative binomial theorem,
\begin{equation}
    (x+y)^{-n}=\sum_{a=0}^\infty (-1)^a\binom{n+a-1}{a}x^{-n-a}y^{a},
\end{equation}
to rewrite $\left(\sin{(2R^2\theta)}\right)^{\frac{\alpha'}{2} k^2}$ as
\begin{equation}
    2i\sum_{a=0}^\infty\binom{-\frac{\alpha'}{2}k^2+a-1}{a}e^{2iR^2\theta\left(\frac{\alpha'}{2}k^2-2a\right)}.
\end{equation}
Note that the binomial coefficient is nothing but a parameter $a$ when applying the on-shell condition.

Unlike the open string case, we choose to consider the limit where  $R\rightarrow 0$ while keeping $\lambda=\alpha'R^2$ fixed. Again, this implies a tensionless string limit, $\alpha'\rightarrow\infty$. In this 'tHooft-like limit, we can rewrite (\ref{closed amp}) as
\begin{align}
    \mathcal{A}(k_1,k_2)=-4\pi^2(2\pi)^d\delta^d(k_1+k_2)g_c^2 \left(\sum_{a=0}^\infty a \right) \int_0^{\infty} d\theta \ e^{-\lambda k^2\theta}
\end{align}
where the Wick rotation: $\theta \rightarrow i\theta$ was used. 
To make sense of the divergent series, we can regulate it via the Riemann zeta function such that 
\begin{equation}
    1+2+3+\ldots=\zeta(-1):=-\frac{1}{12}.
\end{equation}
Thus, the regularized amplitude takes the form
\begin{align}
    \mathcal{A}(k_1,k_2)=\frac{\pi^2}{3}(2\pi)^d\delta^d(k_1+k_2)g_c^2 \int_0^{\infty} d\theta \ e^{-\lambda k^2\theta}
\end{align}
which is exactly the expression (\ref{2pt str2}) upto a constant. Consequently, we can repeat the same computation as the open string amplitude to recast the two-point closed string amplitudes in terms of the AdS transition amplitudes (\ref{relation string and transition amp}), albeit with different prefactors.

Although both open and closed string two-point amplitudes provide the same result, 't Hooft's couplings $\lambda$ are defined differently, where are $\lambda=\alpha'/R$ and  $\tilde\lambda=\alpha'\tilde{R}^{2}$ for open and closed string respectively. $R$ is the radius of a disk while $\tilde{R}$ is the radius of a sphere. It is noted that the factors $\frac{1}{R}$ and $\frac{1}{\tilde{R}^2}$ are the geodesic curvature of the disk and the Gaussian curvature of the sphere. This result might suggest an interesting aspect in open/closed string duality where they may dual through worldsheet curvatures. From our results, an open-string worldsheet with small geodesic curvature gives the same results to that of a closed string with high Gaussian curvature. However, further investigation would be required to establish the full duality.


\section{Conclusions}
We have shown that the two-point amplitudes of open and closed bosonic strings at tree level, when evaluated via the path integral formalism and taken in a tensionless, t’Hooft-like limit, can be reorganized into forms structurally identical to AdS boundary-to-boundary transition amplitudes. This is achieved by expressing both the string amplitudes and AdS transition amplitudes in terms of heat kernel propagators. The two-point string amplitude can be captured as a sum over transition amplitudes presented in (\ref{relation string and transition amp}). 

This may provide further evidence that AdS propagator structures can emerge naturally from both open and closed string amplitudes in flat backgrounds at high-energy ($\alpha'\rightarrow\infty$). Future directions include exploring higher-point functions and examining whether similar AdS structures appear in higher-point string amplitudes.

Additionally, we observe that the ’t Hooft couplings for open and closed string amplitudes depend on different worldsheet curvatures—the geodesic curvature of the disk and the Gaussian curvature of the sphere, respectively. This suggests a potential open/closed string duality through worldsheet geometry, where an open string on a weakly curved disk yields results equivalent to a closed string on a highly curved sphere. Further study is required to fully establish this geometric duality.

\section*{Acknowledgement}
PS is grateful to the National Science, Research and Innovation Fund (NSRF) via the Program Management Unit for Human Resources \& Institutional Development, Research and Innovation for support under grant number B39G680009.



\bibliographystyle{elsarticle-num} 
\bibliography{ref.bib}


\end{document}